\documentclass{JHEP3}

\usepackage{epsfig,multicol,bbm}

\def\mEt{\mbox{${\hbox{$E$\kern-0.6em\lower-.1ex\hbox{/}}}_T$}\, } 

\def\snu{{\tilde\nu}}
\def\snuL{{\tilde\nu_L}}
\def\sN{{\tilde{N}}}
\def\sNR{{\tilde{N}_R}}
\def\snulsp{{\tilde{\nu}_0}}
\def\snuH{{\tilde{\nu}_H}}
\def\stR{{\tilde{t}_R}}

\newcommand{\newc}{\newcommand}

\newc{\mstop}{m_{\tilde{t}}}

\newc{\mtop}{m_t}
\newc{\mbot}{m_b}
\newc{\mz}{m_Z}
\newc{\mw}{M_W}

\newc{\sgn}{\mbox{sgn}}
\newc{\tbeta}{\tan\beta}

\newc{\Mlsp}{M_{\rm LSP}}

\def\beq{\begin{equation}}
\def\eeq{\end{equation}}
\def\bea{\begin{eqnarray}}
\def\eea{\end{eqnarray}}

\preprint{
IFIC/06-01 \\
NUHEP-TH/06-01}

\title{Right-handed Sneutrinos as Nonthermal Dark Matter}

\author{Shrihari Gopalakrishna and Andr\'e de Gouv\^ea \\
Northwestern University, Department of Physics \& Astronomy, 2145 Sheridan Road, Evanston, IL~60208, USA. \\
Email: \email{shri@northwestern.edu, degouvea@northwestern.edu}}

\author{Werner Porod \\
IFIC - Instituto de Fisica Corpuscular, CSIC, E-46071 Valencia, Spain. \\
Email: \email{porod@ific.uv.es} }

\abstract{
When the minimal supersymmetric standard model is augmented by three right-handed neutrino superfields, one generically predicts that the neutrinos acquire Majorana masses. We postulate that all supersymmetry (SUSY) breaking masses as well as the Majorana masses of the right-handed neutrinos are around the electroweak scale and, motivated by the smallness of neutrino masses, assume
that the lightest supersymmetric particle (LSP) is an almost-pure right-handed sneutrino. We discuss the conditions under which this LSP is a successful dark matter candidate. In general, such an LSP has to be nonthermal in order not to overclose the universe, and we find the conditions under which this is indeed the case by comparing the Hubble expansion rate with the rates of the relevant thermalizing processes, including self-annihilation and co-annihilation with other SUSY and 
standard model particles. }


\begin{document}

\section{Introduction}
Neutrino oscillation experiments have revealed that neutrino masses
are nonzero \cite{nu_review}. A renormalizable extension of the standard model (SM)
that incorporates neutrino masses 
can be obtained by introducing right-handed neutrinos
$N_R$ (at least two) which, in conjunction
with the left-handed neutrinos $\nu_L$ of the SM allow neutrino mass
terms. If the masses are assumed to be of the same order of magnitude
as the mass differences, this mass scale is of the order of about
$0.1~$eV. Various proposals have been made for ``explaining'' such a
tiny mass scale, each implying a certain magnitude for the Yukawa
coupling $Y_N$ \cite{nu_theory}.

Supersymmetry (SUSY)~\cite{Martin:1997ns} stabilizes the electroweak scale in which the
quadratic divergences in the Higgs sector due to SM quantum loops are
canceled by contributions from new particles of opposite statistics. In
particular, in the neutrino sector, a supersymmetric extension of the
SM implies the addition (for each generation) of two new complex
scalar fields - the left-handed sneutrino $\snuL$, and the
right-handed sneutrino $\sNR$ - as partners of the left- and
right-handed neutrinos respectively. $N_R$ and $\sNR$ are gauge
singlets under the SM gauge group, and interact with other particles
only through the Yukawa coupling.

It has become evident in recent years through cosmological and
astrophysical observations that there is a significant amount of
non-baryonic dark matter. If R-parity ($R_p$) is conserved, SUSY
provides a stable dark matter candidate - the lightest supersymmetric
particle (LSP)\footnote{In the case of R-parity violation the dark matter
constraint can also be satisfied, e.g.~by a very light and, thus, very
long-lived gravitino LSP \cite{Borgani:1996ag}.}.

If SUSY is realized in nature, it must be broken, and since we do not
know the exact mechanism responsible for SUSY breaking, we will
parametrize this through soft SUSY breaking terms in an effective
Lagrangian at the electroweak scale. SUSY breaking and electroweak
symmetry breaking lead to mixing between the $\snuL$ and $\sNR$, and
we denote the lightest such state as $\snulsp$. We will, henceforth, explore scenarios where
$\snulsp$ is the LSP. In this work
we explore such a dark matter candidate, when all SUSY breaking masses
are of the order of the electroweak scale.

If the $\snulsp$ has a significant component of $\snuL$ which
interacts through the electroweak coupling constants $g$ and
$g^\prime$, the interaction strength is big enough for it to be kept
in thermal equilibrium in the early universe. The final relic
abundance of such a thermal species depends on when it freezes out,
varying inversely as the interaction cross-section. This scenario has
been studied in Refs.~\cite{Hagelin:1984wv,Hall:1997ah,Arkani-Hamed:2000bq,Hooper:2004dc}.
Generally, a predominantly left-handed LSP with mass around 100~GeV
annihilates too efficiently in the early universe resulting in too low a relic-density
today, and reducing the annihilation cross-section by increasing the LSP mass runs 
into conflict with direct detection experiments.
This leaves the possibility of a $\snuL$-$\sNR$ mixed sneutrino as a viable 
dark matter candidate~\cite{Arkani-Hamed:2000bq}.
Another possibility exists, as we will argue here: 
(i) if the $\snulsp$ is almost purely $\sNR$, and 
(ii) if $Y_N$ is sufficiently small, 
the interaction cross-section is too small for the $\snulsp$ to ever be thermalized, 
and the above analyses do not apply.
The question then arises if such a $\snulsp$ can be nonthermal dark matter,
and we investigate in this work under what conditions this can be the case.

The outline of the paper is as follows: In Sec.~\ref{Theory.SEC} we
write down the most general $R_p$ conserving renormalizable superpotential (focusing on
the lepton sector), and the most general renormalizable SUSY breaking
terms, including terms that violate lepton-number. We write down the
sneutrino mass matrix and derive the $\snuL$-$\sNR$ mixing relations.
We impose the constraint from neutrino masses, and specify our
choice of parameters.  In Sec.~\ref{CONDTH.SEC} we write
down the Boltzmann equation for the $\snulsp$ number-density, and
obtain conditions that have to be met in order for the $\snulsp$ to be
thermal. For this we include self-annihilation processes of the
$\snulsp$, annihilation with other SUSY particles, and
annihilation with SM particles. In Sec.~\ref{NonThDM.SEC} we point
out in what cases the above conditions are not met, and thereby
identify when the $\snulsp$ can be nonthermal dark matter.  We offer
our conclusions in Sec.~\ref{CONCL.SEC}.  In App.~\ref{ThRDCalc.APP},
we review in general the standard computation of the relic-density of
a species that was in thermal equilibrium in the early universe, and
apply it to the relic-density calculation of the well-studied mixed
sneutrino case in App.~\ref{ThRelSnu.APP}. Even if the $\snulsp$ is
never in thermal equilibrium, a $\snulsp$ relic-density could result
from the decays of heavier thermal SUSY particles; we compute this
relic-density in App.~\ref{ReThDec.APP}.

\section{The Model}
\label{Theory.SEC}

To the field content of the MSSM, we add (for each generation) the
right-handed neutrino superfield $\widehat N = (\tilde{N}_R, N, F_N)$. Written
as left-chiral fields, the superfields are: $\widehat Q$, $\widehat U^c$, 
$\widehat D^c$, $\widehat L$, $\widehat E^c$, $\widehat N^c$. 
As usual, the MSSM Higgs doublet superfields are $\widehat H_u$
and $\widehat H_d$.

The most general $R_p$ conserving renormalizable superpotential is
\beq
{\cal W} = \widehat U^c Y_U \widehat Q\cdot \widehat H_u
         - \widehat D^c Y_D \widehat Q \cdot \widehat H_d
         + \widehat N^c Y_N \widehat L\cdot \widehat H_u
         - \widehat E^c Y_E \widehat L\cdot \widehat H_d 
		+ \widehat N^c \frac{M_N}{2} \widehat N^c
		+ \mu \widehat H_u \cdot \widehat H_d  \ , 
\label{WSupPot.EQ}
\eeq
where $A \cdot B$ denotes the antisymmetric product of the fields $A$
and $B$, $Y$ are the Yukawa couplings that are $3\times 3$ matrices in
generation space, and, $M_N$ breaks lepton number. In general the
$Y$'s are arbitrary complex matrices, and $M_N$ is complex
symmetric. However, without loss of generality, we can rotate $N^c$ by
a unitary matrix to make $M_N$ real and diagonal with this rotation
matrix absorbed into a redefinition of $Y_N$.

The SUSY breaking terms are
\bea
{\cal L}_{SUSY Br} = &-& \tilde{q}_L^\dagger m^2_{q} \tilde{q}_L 
	- \tilde{u}_R^\dagger m^2_u \tilde{u}_R	- \tilde{d}_R^\dagger m^2_d \tilde{d}_R 
	+ (- \tilde{u}_R^\dagger A_u \tilde{q}_L\cdot h_u 
          + \tilde{d}_R^\dagger A_d \tilde{q}_L\cdot h_d + h.c.) \nonumber \\
	&-& \tilde{\ell}_L^\dagger m^2_{\ell} \tilde{\ell}_L 
	- \tilde{N}_R^\dagger m^2_N \tilde{N}_R	- \tilde{e}_R^\dagger m^2_e \tilde{e}_R 
	+ (- \tilde{N}_R^\dagger A_N \tilde{\ell}_L\cdot h_u
  + \tilde{e}_R^\dagger A_e \tilde{\ell}_L\cdot h_d + h.c.) \nonumber \\
	&+& \left[ (\tilde\ell\cdot h_u)^T \frac{c_{\ell}}{2} (\tilde\ell\cdot h_u) + \tilde{N}_R^T \frac{b_N M_N}{2} \tilde{N}_R + h.c.\right] \nonumber \\
	&+& (b\mu h_u\cdot h_d + h.c.) \ , 
\label{LSUSYBr.EQ}
\eea
where $c_{\ell}$ and $b_N M_N$ break lepton number, and $h_u$ and
$h_d$ are the scalar components of $\widehat H_u$ and $\widehat H_d$ respectively. Note
that $A$ can be an arbitrary complex matrix, and without loss of
generality, $m^2$ is Hermitian while $b_N M_N$ and $c_\ell$ are complex
symmetric.

\subsection{Sneutrino masses}
\label{snuM.SUBSEC}
The sneutrino mass terms resulting from
Eqs.~(\ref{WSupPot.EQ})~and~(\ref{LSUSYBr.EQ}) are (the generation
structure is suppressed)
\bea
{\cal L}^{\tilde\nu}_{mass} &=& -\pmatrix{\snuL^\dagger & \sNR^\dagger & \snuL^T & \sNR^T} {\cal M}_{\snu} \pmatrix{\snuL \cr \sNR \cr \snuL^* \cr \sNR^*} \nonumber \\
{\cal M}_{\snu} &=& \frac{1}{2}
\pmatrix{m_{LL}^2 & m_{RL}^{2\, \dagger} & -v_u^2 c_\ell^\dagger & v_u Y_N^\dagger M_N \cr 
m_{RL}^2 & m_{RR}^2 & v_u M_N^T Y_N^* & - (b_N M_N)^\dagger \cr 
- v_u^2 c_\ell & v_u Y_N^T M_N^* & m_{LL}^{2\, *} & m_{RL}^{2\, T} \cr 
v_u M_N^\dagger Y_N & - b_N M_N & m_{RL}^{2\, *} & m_{RR}^{2\, *}} \ ,
\label{snumassc.EQ}
\eea
where $m_{LL}^2 = (m_\ell^2 + v_u^2 Y_N^\dagger Y_N + \Delta_\nu^2)$, 
$m_{RR}^2 = (M_N M_N^* + m_N^2 + v_u^2 Y_N Y_N^\dagger)$, 
$m_{RL}^2 = (-\mu^* v_d Y_N + v_u A_N)$, and 
$\Delta_\nu^2 = (m_Z^2/2)\cos{2\beta}$ is the D-term contribution. 
Based on the $\snu$ mass matrix, we make the following observations:
\begin{itemize}
\item $m_{RL}^2 = (-\mu^* v_d Y_N + v_u A_N)$ results in $\snuL \leftrightarrow \sNR$ mixing. 
\item $c_\ell$ breaks lepton number and results in $\snuL \leftrightarrow \snuL^*$ mixing~\cite{Hirsch:1997vz,Hall:1997ah}.
\item $M_N$ breaks lepton number and results in $\snuL \leftrightarrow \sNR^*$ mixing. 
\item $b_N M_N$ breaks lepton number and results in $\sNR \leftrightarrow \sNR^*$ mixing.
\end{itemize}

The phases present in Eq.~(\ref{snumassc.EQ}) lead to CP violating
effects which we do not explore in this work.  We can write $\snuL$
and $\sNR$ in terms of real fields $\snu_1$, $\snu_2$, $\sN_1$ and
$\sN_2$ as
\bea
\snuL &=& (\snu_1 + i \snu_2)/\sqrt{2} \ , \nonumber \\
\sNR &=& (\sN_1 + i \sN_2)/\sqrt{2} \ .
\eea
In this basis, if $m_{LL}^2$, $m_{RR}^2$, $m_{RL}^2$, $c_\ell$, $Y_N^\dagger M_N$, and
$b_N M_N$ are all real, Eq.~(\ref{snumassc.EQ}) reduces to block
diagonal form~\cite{Hirsch:1997vz,Grossman:1997is} as
\bea
{\cal L}^{\tilde\nu}_{mass} &=& -\frac{1}{2}\pmatrix{\snu_1^T & \sN_1^T & \snu_2^T & \sN_2^T} {\cal M}_{\snu}^{r} \pmatrix{\snu_1 \cr \sN_1 \cr \snu_2 \cr \sN_2} \nonumber \\
{\cal M}_{\snu}^{r} &=&
\pmatrix{m_{LL}^2 - c_\ell & m_{RL}^{2\, T} + v_u Y_N^T M_N^* & 0 & 0 \cr 
m_{RL}^2 + v_u M_N^\dagger Y_N & m_{RR}^2 - b_N M_N & 0 & 0 \cr 
0 & 0 & m_{LL}^2 + c_\ell & m_{RL}^{2\, T} - v_u Y_N^T M_N^* \cr 
0 & 0 & m_{RL}^2 - v_u M_N^\dagger Y_N & m_{RR}^2 + b_N M_N} \ .
\label{snumassr.EQ}
\eea

We can diagonalize Eq.~(\ref{snumassr.EQ}) by performing the unitary rotation
\bea
\pmatrix{\snu_i \cr \sN_i} &=& \pmatrix{\cos{\theta^{\snu}_i} & -\sin{\theta^{\snu}_i} \cr \sin{\theta^{\snu}_i} &
 \cos{\theta^{\snu}_i}} \pmatrix{\snu_i^\prime \cr \sN_i^\prime} \nonumber \\
         &\equiv& \left( {\cal C}^{\snu}_i \right) \pmatrix{\snu_i^\prime \cr \sN_i^\prime} \ ,
\label{snumixang.EQ}
\eea
where the mixing angle is given by
\bea
\tan{2\theta^{\snu}_i} &=& \frac{2\left( m_{RL}^2 \pm v_u M_N^\dagger Y_N \right)}{(m_{LL}^2 \mp c_\ell) - (m_{RR}^2 \mp b_N M_N)} \ , \label{thsnu.EQ}
\eea
with the top (bottom) sign for $i=1$ ($i=2$). We will use the notation: 
$s_i \equiv \sin(\theta^{\snu}_i)$, $c_i \equiv \cos(\theta^{\snu}_i)$ and 
$t_i \equiv \tan(\theta^{\snu}_i)$.

We see from Eq.~(\ref{snumassr.EQ}) that the lepton-number-violating
parameters $c_\ell$, $b_N$ and $M_N$ split the $\snu_1 \leftrightarrow
\snu_2$ degeneracy, and also the $\sN_1 \leftrightarrow \sN_2$
degeneracy. We denote the lightest of the sneutrino mass eigenstates
as $\snulsp$, and its mass $\Mlsp$. We will also assume that $\snulsp$
is the lightest supersymmetric particle (LSP), with $R_p$ conserved,
so that it is a stable particle. For definiteness, we will assume that
the LSP is in the upper-left $2\times 2$ block. We denote the heavier
sneutrino mass eigenstates collectively as $\snuH$.

The hidden generation structure in Eq.~(\ref{snumassr.EQ}) in general
implies new flavor changing neutral current (FCNC) contributions,
which are experimentally constrained to be small \cite{pdg}.  These contributions are
small for instance if the fermion and scalar mass matrices are
suitably aligned. Moreover, an alignment at the high (GUT) scale is
sufficient, since we expect that renormalization-group loop-induced
violations of this alignment to be proportional to $Y_N$, which, as will be discussed
shortly, we assume to be very small.

\subsection{Neutrino masses}
The superpotential in Eq.~(\ref{WSupPot.EQ}) leads to the neutrino mass terms
($N^c$ and $\nu$ are 2-component spinors)
\bea
{\cal L}^{\nu}_{mass} = - N^c v_u Y_N \nu
 - N^c \frac{M_N}{2} N^c + h.c. \ .
\eea
When $v_u Y_N \ll M_N$, the lowest neutrino mass eigenvalue 
is given by the standard seesaw relation
\beq
m_{\nu} = \frac{v_u^2 Y_N^2}{M_N} \ .
\label{mnuss.EQ}
\eeq

\subsection{Choice of parameters}
The masses of the light neutrinos, $m_{\nu}$, are required to be less than about $0.1~{\rm eV}$ to 
correspond to the mass scale inferred from neutrino oscillation experiments.
The usual Type I seesaw results if $M_N \sim 10^{14}~$GeV and $Y_N \sim O(1)$. 
However, this is not the only possibility; $M_N$ could be at the 
electroweak scale~\cite{Arkani-Hamed:2000bq,Hooper:2004dc,Borzumati:2000mc}, 
or even at the eV scale~\cite{deGouvea:2005er}. 
If lepton number is a good symmetry (i.e., if $M_N = b_N = c_\ell = 0$), a Dirac neutrino
results, and $Y_N \sim 10^{-12}$ in order to have $m_\nu \sim 0.1~{\rm eV}$. 

We will parameterize $A_N$ such that
\beq
A_N = a_N Y_N m_\ell  \ ,
\label{natAN.EQ}
\eeq 
where $a_N$ is a dimensionless constant. We further assume $a_N$ to be order one, {\it i.e.}, if the neutrino Yukawa couplings turn out to be very small (which is the case here), the neutrino $A_N$ also turn out to be significantly smaller than the weak scale. This assumption is automatically satisfied in the constrained MSSM \cite{pdg} (sometimes referred to as ``minimal supergravity'' MSSM). Other SUSY breaking mechanism that yield zero A-terms at a low-enough SUSY-breaking scale would also qualify.
In this case, Eq.~(\ref{thsnu.EQ}) then implies that 
\beq
s_1 \approx Y_N \frac{v_u}{m_\ell} \alpha_m \ ; \quad 
\alpha_m \equiv \left(a_N + \frac{M_N - \mu^* \cot\beta}{m_\ell} \right)  \ .
\label{nats1.EQ}
\eeq
In this work we consider the case when all other relevant mass-scales are at
the electroweak scale, including the neutrino Majorana mass ($M_N$).
In order to obtain $m_\nu \sim 0.1~{\rm eV}$, we see from
Eq.~(\ref{mnuss.EQ}) that we require $Y_N \sim 10^{-6}$. We take the mass of $\sNR$ to
also be around the electroweak scale, i.e., $(m_{RR}^2 - b_N M_N) \sim
v$ (cf. Eq.~(\ref{snumassr.EQ})).  For $Y_N \sim 10^{-6}$ and $v_u,
m_\ell \sim 10^2~$GeV, from Eqs.~(\ref{natAN.EQ}) and
(\ref{nats1.EQ}), we have:
\beq
Y_N \sim 10^{-6} \ ;  \quad A_N \sim a_N \cdot 100~{\rm keV} \ ;\quad s_1 \sim 10^{-6} \alpha_m \ .
\label{natnums.EQ}
\eeq 
Note that, as long as $a_N$ is indeed order one (as we assume above), $\alpha_N$ is also expected to be of order one (see Eq.~\ref{nats1.EQ}), and hence $s_1$ is guaranteed to be very small. Hence, we will be interested in exploring such a predominantly right-handed LSP ($\snulsp \approx \sNR$)
as a dark matter candidate, with $s_1 \ll 1$ and $c_1 \approx 1$.

\section{When is the $\snulsp$ thermal?}
\label{CONDTH.SEC}
Depending on the $\snulsp$ interaction strength with itself and other particles, it can 
either thermalize if the interaction rate is bigger than the Hubble expansion rate, 
or remain nonthermal if the interactions are too weak. In this section we identify what 
conditions have to be consistently satisfied for $\snulsp$ to be in thermal equilibrium 
with the SM thermal bath at some moment in the thermal history of the universe.

\subsection{Boltzmann equation}
The $\snulsp$ number density $n_{\snulsp}$ is governed by the Boltzmann equation
\beq
\frac{d}{dt} n_{\snulsp} = - 3 H n_{\snulsp} - \left< \sigma v \right>_{SA} 
\left(n_{\snulsp}^2 - n_{\snulsp\,eq}^2 \right) - \left< \sigma v \right>_{CA} 
\left(n_{\snulsp} n_{\phi} - n_{\snulsp\,eq} n_{\phi\,eq}  \right)
+ C_\Gamma \ ,
\label{sNBoltz.EQ}
\eeq
where the subscripts $SA$ and $CA$ on the thermally averaged cross-section 
$\left<\sigma v \right>$ denote self-annihilation and co-annihilation (annihilation with another
species $\phi$) respectively,
$n_{eq}$ denotes the equilibrium number density, and $C_\Gamma$ is the contribution
due to decay of heavier particles into $\snulsp$. 

$C_\Gamma$ includes the contribution from all heavier SUSY particles that can decay into $\snulsp$. 
In App.~\ref{ReThDec.APP}, we identify such heavier SUSY particles, 
and, assuming they are in thermal equilibrium, compute the present 
relic density of the $\snulsp$ by integrating the Boltzmann equation. It turns out that these 
contributions could contribute too big of a $\snulsp$ relic-density, and have to be 
``dealt with'' appropriately.
In the remainder of this section we will consider the effects of the other terms in the Boltzmann equation.
 
The $\snulsp$ number density will be driven to its equilibrium value provided that the interaction
rate is bigger than the Hubble expansion rate, {\it i.e.}, if 
\beq
\left< \sigma v \right>_{SA} n_{\snulsp}  > 3 H \ ; \qquad
\left< \sigma v \right>_{CA} n_{\phi}  > 3 H \ ,
\eeq
for the self- and co-annihilation channels respectively.  
For annihilation of non-relativistic scalar particles ($\snulsp$ here), we can 
make the Taylor expansion~\cite{Jungman:1995df}
$\left< \sigma v \right> = a + b v^2 + \cdots \approx a\, +\, 6\, b\, T/M$,
where the first (second) term is the s-wave (p-wave) contribution. 
Using Eq.~(\ref{HT2.EQ}) for $H$, we get
\beq
\left<\sigma v\right>_{SA} n_{\snulsp}  > 3 \left( 1.66 g_*^{0.5} \frac{T^2}{M_{Pl}} \right) \ ; \qquad
\left<\sigma v\right>_{CA} n_{\phi}  > 3 \left( 1.66 g_*^{0.5} \frac{T^2}{M_{Pl}} \right) \ , 
\label{ThCond1.EQ}
\eeq
as the condition for thermal equilibrium for the self- and co-annihilation channels respectively.

Next, we check for various processes whether the inequality Eq.~(\ref{ThCond1.EQ}) is satisfied 
for $T\approx \Mlsp$; if we find that equilibrium is not established for $T\approx \Mlsp$, it
will not be for $T \gg \Mlsp$. This is because the cross-section at most
goes like $\sigma \propto 1/T^2$, while the number density goes like $n \propto T^3$,  
which implies that the left-hand-side of 
Eq.~(\ref{ThCond1.EQ}) goes like $T$ while the right-hand-side goes like $T^2$.
Therefore, if the inequality is not satisfied at $T \approx \Mlsp$ it will 
never be satisfied for $T \gg \Mlsp$. For this reason we only need to check if the condition
is satisfied for $T \lesssim \Mlsp$.

\subsection{Self-annihilation}
\label{SELFINT.SUBSEC}
The dominant $\snulsp$ self-annihilation processes are\footnote{
The process, $\snulsp \snulsp \rightarrow \psi \overline{\psi}$ (where $\psi$ is a SM fermion) 
due to s-channel $Z$-boson exchange, is not present since the coupling of a pair of identical scalars
to the $Z$-boson is zero.}:
\begin{itemize}
   \item[($a_s$)] $\snulsp \snulsp \rightarrow \nu_L \nu_L$ via t-channel exchange of (Majorana) 
$\tilde W^3$ and $\tilde B$.
   \item[($b_s$)] $\snulsp \snulsp \rightarrow Z Z\ ;\ W^+ W^-$ via t-channel exchange of 
$\snuH ;\ \tilde\ell$. 
   \item[($c_s$)] $\snulsp \snulsp \rightarrow \psi \overline{\psi}$ (where $\psi=(c,t,b)$) via 
s-channel exchange of $h_u\ ;\ h_d$.\footnote{ Associated with channels involving $h_u, h_d$ 
exchange, there are analogous channels with a longitudinal gauge-boson (Goldstone boson) 
exchange. We do not explicitly show these since their contributions are of the same order of 
magnitude as the $h_u, h_d$ contributions discussed.}
   \item[($d_s$)] $\snulsp \snulsp \rightarrow \nu_L \overline{\nu_L}\ ;\ e_L \overline{e_L}$ via 
t-channel exchange of $\tilde H_u^+\ ;\ \tilde H_u^0$. 
\end{itemize}

In Table~\ref{SelfInt.TAB} we show the cross-sections for the above processes for 
$T \lesssim \Mlsp$, under the simplifying assumption that the mass of the exchanged particle 
is much bigger than $T$. $Y_\psi$ is the Yukawa coupling of $\psi$.
The limit shown in the table is got from the inequality Eq.~(\ref{ThCond1.EQ}), and if the 
limit is satisfied, that particular process can thermalize $\snulsp$. The inequality is evaluated 
at $T \sim \Mlsp$, with $n_{\snulsp} \sim T^3 \approx \Mlsp^3$, with all SUSY masses taken to be 
around 100~GeV, and for $c_1 \approx 1$.
In process $(d_s)$, we include the factor $(T/\Mlsp)$ due to the p-wave annihilation, 
and ignore the s-wave contribution that is helicity-suppressed by 
the factor $(m_e/\Mlsp)^2 \approx 10^{-4}$ (for the $\tau$ lepton). 
In deriving the limit for process $(c_s)$ we assume $M_{h_u} \approx M_{h_d} \equiv M_h$, 
and make use of the definition $\alpha_N \equiv (a_N - \mu^*/m_\ell)$. For large $\tan\beta$, the
bottom final-state can be important since $Y_\psi = Y_b$ can be $O(1)$.
$f_{PS}$ denotes the phase-space factor which can be a severe suppression for the top final-state.

\TABLE[t]{
\caption{Self-annihilation cross-sections and limits. 
We convert the limit in the third column to an implied limit shown in the last column using
Eqs.~(\ref{natAN.EQ}) and (\ref{nats1.EQ}).
 \label{SelfInt.TAB} } 
\vspace*{.2cm}
\begin{centering}\begin{tabular}{|c|c|c|c|}
\hline 
Process & Cross-section & Limit & Implied limit \tabularnewline
\hline
\hline 
$(a_s)$ &  
$\frac{s_1^4}{16\pi} \left(\frac{g^2}{M_{\tilde W}}+ \frac{{g^\prime}^2}{M_{\tilde B}} \right)^2 $
  & $s_1 > 10^{-3}$ & $\alpha_m Y_N > 10^{-3}$ \tabularnewline
\hline 
$(b_s)$ & $\frac{s_1^4}{16\pi} \left(\frac{g^2 + {g^\prime}^2}{M_{\snuH}^2} + \frac{g^2 + {g^\prime}^2}{M_{\tilde e_L}^2}\right)^2 \Mlsp^2$ & $s_1 > 10^{-2}$ & $\alpha_m Y_N > 10^{-2}$  
\tabularnewline
\hline 
$(c_s)$ & $\frac{s_1^2 c_1^2 Y_\psi^2}{16\pi} \frac{|A_N-\mu^* Y_N|^2}{M_{h}^4} f_{PS} $  &  
$Y_\psi s_1 |A_N-\mu^* Y_N| f_{PS} > 10~{\rm keV}$ & $Y_\psi \alpha_N \alpha_m Y_N f_{PS} > 10^{-2} $  
\tabularnewline
\hline 
$(d_s)$ & $\frac{Y_N^4 c_1^4}{16\pi} \frac{1}{M_{\tilde H}^2} \frac{T}{\Mlsp} $ 
& $Y_N > 10^{-3.5}$ & $Y_N > 10^{-3.5}$ \tabularnewline
\hline
\end{tabular}\par\end{centering}
}

For processes that limit parameters other than $Y_N$, we show in the last column of 
Table~\ref{SelfInt.TAB} 
implied limits on $Y_N$, assuming Eqs.~(\ref{natAN.EQ}) and (\ref{nats1.EQ}).
Among all the self-annihilation processes, $(d_s)$ leads to the strongest limit 
$Y_N > 10^{-3.5}$.
Therefore, for the case of interest, Eq.~(\ref{natnums.EQ}), we infer that self-annihilation 
processes are not effective in thermalizing $\snulsp$.

\subsection{Co-annihilation with SUSY particles}
\label{COINT.SUBSEC}
The $\snulsp$ can thermalize by co-annihilation with other SUSY particles through processes such as
($\psi$ denotes a SM fermion, and $\tilde s$ denotes SUSY particles other than $\tilde\nu$):
\begin{itemize}
   \item[$(a_c)$] $\snulsp \tilde \nu_H  \rightarrow \psi \overline{\psi}$  via s-channel exchange of $Z$-boson. 
   \item[$(b_c)$] $\snulsp \tilde{s} \rightarrow \tilde{e}_L \tilde{s}^\prime$ via t-channel 
exchange of $W^{\pm}$. 
   \item[$(c_c)$] $\snulsp \tilde t_R^* \rightarrow e_R \overline{b_L}$ via t-channel exchange of 
$\tilde H^\pm$.
   \item[$(d_c)$] $\snulsp \tilde t_R^* \rightarrow e_L \overline{b_L}\ ;\ \nu_L \overline{t_L}$ 
via t-channel exchange of $\tilde H_u^+\ ;\ \tilde H_u^0$.
   \item[$(e_c)$] $\snulsp \snuH \rightarrow \psi \overline{\psi}$ (where $\psi =(c,t,b)$) 
via s-channel exchange of $h_u ;\ h_d$.
\end{itemize}
We recall here that if a heavier species $\phi$ is in thermal equilibrium, its number 
density at a temperature $T\sim \Mlsp$ is Boltzmann suppressed (cf. Eq.~(\ref{nBolsupr.EQ})) 
compared to the thermal number density of $\snulsp$
by the factor $\zeta_\phi \equiv e^{-(\Delta M_\phi/T)}$, where 
$\Delta M_\phi \equiv (M_\phi - \Mlsp)$.

In Table~\ref{CoIntSUSY.TAB} we show cross-sections and limits that arise from the above
co-annihilation processes with heavier SUSY particles. We denote the Boltzmann suppression of
the heavier species $\phi$ as $\zeta_\phi$ (see above), the phase-space suppression factor due to 
heavy particles in the final state as $f_{PS}$, and the $\psi$ Yukawa coupling as $Y_\psi$.
Processes $(a_c)$ and $(d_c)$, like $(d_s)$ in the previous subsection, has the p-wave annihilation 
factor $(T/\Mlsp)$, and we ignore the helicity-suppressed s-wave contribution. 
The implied limit if Eqs.~(\ref{natAN.EQ}) and (\ref{nats1.EQ}) hold is shown in the last column.
\TABLE[t]{
\caption{Co-annihilation with SUSY - cross-sections and limits. 
We convert the limit in the third column to an implied limit shown in the last column using
Eqs.~(\ref{natAN.EQ}) and (\ref{nats1.EQ}).
\label{CoIntSUSY.TAB}}
\vspace*{.2cm}
\begin{centering}\begin{tabular}{|c|c|c|c|}
\hline 
Process & Cross-section & Limit & Implied limit \tabularnewline
\hline
\hline 
$(a_c)$  &  $\frac{\left(g^2 + g^{\prime\, 2} \right)^2 s_1^2 c_1^2}{16\pi} \frac{1}{M_Z^2} \frac{T}{\Mlsp}$
  &  $\zeta_{\snuH} s_1 > 10^{-6.5}$ & $\zeta_{\snuH} \alpha_m Y_N > 10^{-6.5} $ \tabularnewline
\hline 
$(b_c)$  &  $\frac{g^4 s_1^2}{16\pi} \frac{\Mlsp^2}{M_Z^4} f_{PS}^2$  &  
$\zeta_{\tilde s} s_1 f_{PS} > 10^{-6.5}$ & $\zeta_{\tilde s}\alpha_m Y_N f_{PS} > 10^{-6.5}$ \tabularnewline
\hline 
$(c_c)$  &  $\frac{Y_e^2 Y_t^2 s_1^2}{16\pi} \frac{\mu^2}{M_{\tilde H}^4}$  &  
$\zeta_{\stR} s_1 > 10^{-5}$ & $\zeta_{\stR}\alpha_m Y_N > 10^{-5}$ \tabularnewline
\hline 
$(d_c)$  &  $\frac{Y_N^2 c_1^2 Y_t^2}{16\pi} \frac{1}{M_{\tilde H}^2} \frac{T}{\Mlsp}$
&  $\zeta_{\stR} Y_N > 10^{-6.5}$  &  $\zeta_{\stR} Y_N > 10^{-6.5}$
 \tabularnewline
\hline
$(e_c)$  &  $\frac{(c_1^2 - s_1^2)^2 Y_\psi^2}{16\pi} \frac{|A_N - \mu^* Y_N|^2}{M_h^4} f_{PS}^2$  
&  $\zeta_{\snuH} Y_\psi |A_N -\mu^* Y_N| f_{PS} > 10~$keV  &  
$Y_\psi \zeta_{\snuH} \alpha_N Y_N f_{PS} > 10^{-7} $ \tabularnewline
\hline
\end{tabular}\par\end{centering}
}
We have defined, as before, $\alpha_N \equiv (a_N - \mu^*/m_\ell)$.

As we see from Table~\ref{CoIntSUSY.TAB}, though the processes impose strong constraints, 
if phase-space and Boltzmann suppression factors are strong enough, 
for the case of interest here (Eq.~(\ref{natnums.EQ})), we expect co-annihilation processes
with heavier SUSY particles not to be effective in thermalizing $\snulsp$.
We note that if $\tan(\beta)$ is large, co-annihilation with $\tilde b_R$ can also be important, 
and for $Y_b \approx 1$ will lead to limits similar to that from processes $(c_c)$ and $(d_c)$. 
Also, in this case, in process $(e_c)$ the bottom final state is important.

\subsection{Co-annihilation with SM particles}
\label{SMINT.SUBSEC}
The $\snulsp$ can thermalize through interactions with thermal 
populations of SM particles. The important processes are ($\psi$ denotes a SM particle):
\begin{itemize}
   \item[$(a_M)$] $\snulsp \psi_{R,L} \rightarrow \tilde{\nu}_H \psi_{L,R}$ (where $\psi=(c,t,b)$) 
via t-channel $h_u ;\ h_d$ exchange.
   \item[$(b_M)$] $\snulsp \psi_{L} \rightarrow \nu_L \tilde{\psi}_{R}$ (where $\psi=(c,t,b)$) 
via t-channel $\tilde H$ exchange.
   \item[$(c_M)$] $\snulsp \psi \rightarrow  \tilde{e}_L \psi^\prime$ via t-channel $W^\pm$ exchange.
   \item[$(d_M)$] $\snulsp \psi \rightarrow  \snuH \psi$ via t-channel $Z$ exchange.
   \item[$(e_M)$] $\snulsp \psi \rightarrow L \tilde{\psi}$ via t-channel $\tilde{W}$ exchange. 
\end{itemize}

Table~\ref{CoIntSM.TAB} shows the cross-sections and the limits due to co-annihilation of 
$\snulsp$ with SM particles. 
As before, in process $(a_M)$ we define $\alpha_N \equiv (a_N - \mu^*/m_\ell)$, and we compute
the limit assuming $M_{h_u} \approx M_{h_d} \equiv M_h$.
In $(e_M)$, the $\tilde{B}$ channel can also contribute but is less important given the smaller
$g^\prime$.

\TABLE[t]{
\caption{Co-annihilation with SM - cross-sections and limits. 
We convert the limit in the third column to an implied limit shown in the last column using
Eqs.~(\ref{natAN.EQ}) and (\ref{nats1.EQ}).
\label{CoIntSM.TAB}} 
\vspace*{.2cm}
\begin{centering}\begin{tabular}{|c|c|c|c|}
\hline 
Process & Cross-section & Limit & Implied limit \tabularnewline
\hline
\hline 
$(a_M)$  &  $\frac{Y_\psi^2 (c_1^2-s_1^2)^2}{16\pi} \frac{|A_N -\mu^* Y_N|^2}{M_{h}^4} f_{PS}^2 $
& $Y_\psi \zeta_\psi |A_N -\mu^* Y_N| f_{PS} > 10~$keV  &  $Y_\psi \zeta_\psi \alpha_N Y_N f_{PS} > 10^{-7} $ \tabularnewline
\hline 
$(b_M)$  &  $\frac{Y_N^2 Y_\psi^2 c_1^2}{16\pi} \frac{1}{M_{\tilde H}^2} f_{PS}^2 $  &  
$Y_\psi \zeta_\psi Y_N f_{PS} > 10^{-7} $  &  $ Y_\psi \zeta_\psi Y_N f_{PS} > 10^{-7} $ 
\tabularnewline
\hline 
$(c_M)$  &  $\frac{g^4 s_1^2}{16\pi} \frac{\Mlsp^2}{M_W^4} f_{PS}^2 $  &  $s_1 f_{PS} > 10^{-6.5} $  &  $\alpha_m Y_N f_{PS} > 10^{-6.5} $ \tabularnewline
\hline 
$(d_M)$  &  $\frac{g^4 s_1^2 c_2^2}{16\pi} \frac{\Mlsp^2}{M_Z^4} f_{PS}^2 $  &  $s_1 f_{PS} > 10^{-6.5} $  &  $\alpha_m Y_N f_{PS} > 10^{-6.5} $ \tabularnewline
\hline
$(e_M)$  &  $\frac{g^4 s_1^2}{16\pi} \frac{\Mlsp^2}{M_{\tilde W}^2} f_{PS}^2 $  &  $s_1 f_{PS} > 10^{-6.5} $  &  $\alpha_m Y_N f_{PS} > 10^{-6.5} $ \tabularnewline
\hline
\end{tabular}\par\end{centering}
}

In $(a_M)$ and $(b_M)$, the charm and bottom channels are suppressed by the smaller Yukawa 
couplings compared to the top, and for $T < 100~$GeV the top contribution is Boltzmann suppressed. 
In general, processes $(a_M)$ and $(c_M)$-$(e_M)$ are suppressed further due to $f_{PS} \ll 1$ 
if the left-handed SUSY particles in the final-state are all much heavier.
If $\tan\beta$ is large, co-annihilation with bottom, especially $(b_M)$, will lead to the strongest 
constraint since it is not Boltzmann suppressed (unless $T < m_b$). 
However, $(b_M)$ may be suppressed if $M_{\tilde H} \gg 100~$GeV.
Thus a combination of suppressions from $f_{PS}$, $\zeta_\psi$ and $M_{\tilde{W},\tilde{H}}$ can
render these processes incapable of thermalizing $\snulsp$.

\section{Nonthermal $\snulsp$}
\label{NonThDM.SEC}
If the $\snulsp$ is never in thermal equilibrium, the standard relic density calculation 
presented in App.~\ref{ThRDCalc.APP} cannot be applied. 
We consider in this section such a situation when an almost purely right-handed LSP 
($\snulsp \approx \sNR$) is not in thermal equilibrium, and the possibility of it being 
nonthermal dark matter. 
In order for such a situation to be realized, it should be true that all the $\snulsp$ 
thermalization conditions that we discussed in Sec.~\ref{CONDTH.SEC} are not satisfied.
For the case specified in Eq.~(\ref{natnums.EQ}),
this happens for instance in the inflationary paradigm for a low reheat 
temperature ($T_{RH} \ll 100~$GeV), whereby the reheating produces only $\sNR$ (and SM particles),
and the number-densities of heavier SUSY particles (and top) are severely Boltzmann suppressed. 
We will elaborate on this in the following.

First, as we found in Sec.~\ref{SELFINT.SUBSEC}, none of the self-annihilation processes
can keep the $\snulsp$ in thermal equilibrium for the parameter values in Eq.~(\ref{natnums.EQ}).

Second, from Sec.~\ref{COINT.SUBSEC}, the co-annihilation processes can be 
important in thermalizing the $\snulsp$ for the parameter values in Eq.~(\ref{natnums.EQ}). 
However, $(b_c)$ is suppressed further due to 
$f_{PS} \ll 1$ which we expect given the heavy particles in the final state, 
$(d_c)$ is p-wave suppressed for $T_{RH} \ll \Mlsp$, and, $(a_c)$ and $(e_c)$ 
are Boltzmann suppressed due to the heavy $\snuH$; hence we expect these 
processes also to be ineffective in thermalizing $\snulsp$.

Third, the processes of Sec.~\ref{SMINT.SUBSEC} can all be important.
However, phase-space suppression $f_{PS} \ll 1$ due to
the heavy particles in the final state, the Boltzmann suppression due to the heavy top 
(for $T_{RH} < M_t$), and suppression due to $M_{\tilde{W},\tilde{H}} > 100~$GeV can render
these processes ineffective in thermalizing the $\snulsp$.
 
Fourth, for the parameter values in Eq.~(\ref{natnums.EQ}), the $\snulsp$ relic-density from decays of 
heavier SUSY thermal particles are problematic and tend to overclose the universe if thermal 
populations of heavier SUSY particles are present (cf. App.~\ref{ReThDec.APP}). 
However, for $T_{RH} \ll 100~$GeV, with no significant number-densities of heavier 
SUSY particles, this problem is avoided.

If such a situation of a nonthermal $\sNR$ is realized, the relic abundance is determined by
how the $\sNR$ is produced in the early universe. For example, if the universe went 
through an inflationary epoch driven by a scalar field $\Phi$ (the inflaton), 
the relic abundance today becomes directly related to the coupling of $\Phi$
to the $\sNR$ (and the other particles), and the details of reheating.

\section{Conclusions and future directions}
\label{CONCL.SEC}
We analyzed the sneutrino sector after writing down the most general $R_p$ conserving 
renormalizable effective theory, including lepton-number violating terms. We considered the case 
when the neutrino Majorana mass term $M_N$, and all SUSY breaking masses are at the electroweak 
scale, which implied that the neutrino Yukawa coupling $Y_N \sim 10^{-6}$. 
Assuming that the A-term is proportional to $Y_N$ we were led to a tiny sneutrino mixing angle
$s \sim Y_N$. We analyzed the cosmological implications of having such a predominantly 
right-handed sneutrino LSP ($\snulsp \approx \sNR$), for values given in Eq.~(\ref{natnums.EQ}).

Since such a $\snulsp$ interacts mostly through the tiny $Y_N$, its self- and co-annihilation 
cross-sections are tiny. Therefore, the $\snulsp$ should not be in thermal equilibrium at any
time in the early universe, for if it is, the standard thermal relic-density calculation 
(reviewed in Apps.~\ref{ThRDCalc.APP} and \ref{ThRelSnu.APP}) indicates that it would 
severely overclose the universe since it freezes-out too early.
Moreover, we argue in App.~\ref{ReThDec.APP} that decays into $\snulsp$ from heavier thermal 
SUSY particles, for these parameters, also overcloses the universe (see also \cite{Asaka:2005cn}). 

The main focus of the paper is in investigating in which case the $\snulsp$ remains nonthermal,
so that the problems mentioned in the previous paragraph for a thermal relic do not apply.
In order to answer this, we start with the Boltzmann equation for the $\snulsp$ number-density,
and compare its self- and co-annihilations with the Hubble expansion rate; as shown in 
Eq.~(\ref{ThCond1.EQ}). For a given process,
if the interaction cross-section is smaller than the Hubble rate, this process
cannot thermalize the $\snulsp$. We check to see whether dominant interaction processes of the 
$\snulsp$ with itself, other SUSY particles, and SM particles, satisfy this condition.  

In Sec.~\ref{SELFINT.SUBSEC} we find from Table~\ref{SelfInt.TAB} that the self-annihilation
processes are ineffective in thermalizing such a $\snulsp$ for the values in 
Eq.~(\ref{natnums.EQ}). 
Next, in Sec.~\ref{COINT.SUBSEC}, Table~\ref{CoIntSUSY.TAB} we find that $\snulsp$ 
co-annihilation processes with other SUSY particles (if they are present in thermal 
number-densities)
are quite effective in thermalizing the $\snulsp$. However including phase-space and
Boltzmann suppression factors shown in the table can render these processes ineffective.
Lastly, in Sec.~\ref{SMINT.SUBSEC}, Table~\ref{CoIntSM.TAB} we find that co-annihilation
processes with SM particles are significant in thermalizing the $\snulsp$; however, the phase-space,
Boltzmann, and heavy intermediate particle mass suppression factors can render them ineffective.

In Sec.~\ref{NonThDM.SEC} we argued that in the inflationary paradigm, if the reheat temperature
$T_{RH} \ll 100~$GeV, all of these constraints are rendered harmless and the $\sNR$ is nonthermal.
The relic number density will then be connected to details 
of reheating and the $\snulsp$ coupling to the inflaton. Exploring these details and of 
how baryogenesis works in this scenario is, however, beyond the scope of this 
work~\cite{OtherPapers}. 

In colliders, any heavy SUSY particle that is produced undergoes
cascade decays into the $\snulsp$ LSP, which exits the detector as
missing energy. An important signature associated with a predominantly $\sNR$ LSP is
the occurrence of  a displaced vertex in the detector in the case of Majorana 
neutrinos, or even a very long-lived charged particle leaving the detector
in the case of Dirac neutrinos as the life-time
scales like $|Y_N|^{-2}$. Further details will be discussed in an
up-coming paper~\cite{OurCollPaper}.

In summary, if all new physics scales, including the SUSY breaking and the lepton number breaking scales are small, there is a good chance that the LSP is the lightest sneutrino $\snulsp$. Furthermore, depending on the mechanism of SUSY breaking, the $\snulsp$ could naturally be an almost-pure right-handed sneutrino. Here, we argue that, if the $\snulsp$ is stable, such a scenario is not only cosmologically allowed, but can also properly fit our understanding of dark matter, as long as very stringent requirements on ``initial conditions'' are met. 
Such a scenario may also address some of the problems with the standard cold dark matter
(CDM) paradigm~\cite{CDMprobsPapers}.
Moreover, collider signatures and  signatures associated with such a dark matter candidate seem unusual enough to render such a scenario phenomenologically intriguing.

\vspace*{5mm}
\noindent
{\bf Acknowledgments}

\noindent 
We  thank the organizers and participants of the LC workshop
Snowmass 2005 where this work was
initiated. We thank Csaba Balazs, Marcela Carena, Martin Hirsch, Dan Hooper and
James Wells for valuable discussions. The work of AdG and SG is sponsored in part by the US Department of Energy Contract DE-FG02-91ER40684, while WP is supported by a 
MEC Ramon y Cajal contract and by the Spanish grant BFM2002-00345.

\appendix
\section{Thermal relic density calculation}
\label{ThRDCalc.APP}
Observations indicate at a high confidence level that the universe is flat, and we therefore assume
that such is the case in presenting all the formulas here. 
In order to derive the relic abundance of the dark matter candidate $\chi$,
we start with the Friedmann equation for a flat universe
\beq
H^2 = \frac{8\pi G}{3} \rho \ , 
\eeq
which can be written during the radiation-dominated era in terms of the temperature as
\beq
H(T) = 1.66 g_*^{0.5} \frac{T^2}{M_{Pl}} \ ,
\label{HT2.EQ}
\eeq
where $g_*$ is the effective number of relativistic degrees of freedom at $T$.
The species freezes out at the temperature $T_f$ when the self-annihilation rate of $\chi$
becomes roughly equal to the Hubble expansion rate $H$, i.e.,
\beq
\left< \sigma v\right> n_\chi \sim H(T_f) \ ,  
\label{eqIntH.EQ}
\eeq
where $\left< \sigma v\right>$ is the thermally averaged cross-section and
$n$ is the number density of $\chi$. 
The number density is given as~\cite{CosmoBooks}
\bea
n(T) = \left\{
	   \begin{array}{c}
	       (\zeta(3)/\pi^2) g T^3 \qquad\quad\quad {\rm (Boson)}, \ T \gg M \ , \\
    	       (3/4)(\zeta(3)/\pi^2) g T^3 \quad\quad {\rm (Fermion)}, \ T \gg M \ , \\
     	       g\left(\frac{MT}{2\pi}\right)^{3/2} e^{-(M-\mu)/T} \qquad\qquad\quad T \sim M \ , 
	   \end{array}
       \right.		
\label{numdenT.EQ}
\eea
where $g$ is the number of degrees of freedom of $\chi$ (for example, $g=1$ for a real scalar,
$g=2$ for a Dirac fermion, etc.).

If $\chi$ can co-annihilate with another species $\phi$ then we can generalize 
Eq.~(\ref{eqIntH.EQ}) to
\beq
\left< \sigma v\right>_{SA} n_\chi + \left< \sigma v\right>_{CA} n_\phi \sim H(T_f) \ ,  
\label{eqCoIntH.EQ}
\eeq
where the subscripts $SA$ and $CA$ on the cross-sections denote self- and co-annihilation 
respectively. When $\chi$ and $\phi$ are in thermal equilibrium, and furthermore if
$M_\phi > M_\chi$, Eq.~(\ref{numdenT.EQ}) allows us to write
\beq
n_\phi = \left(\frac{M_\phi}{M_\chi}\right)^{(3/2)} e^{-(\Delta M_\phi/T)} n_\chi \qquad\qquad T\sim M_\chi \ ,
\label{nBolsupr.EQ}
\eeq
with $\Delta M_\phi \equiv (M_\phi - M_\chi)$. Then Eq.~(\ref{eqCoIntH.EQ}) becomes
\beq
\left[ \left< \sigma v\right>_{SA}  + \left< \sigma v\right>_{CA} \left(\frac{M_\phi}{M_\chi}\right)^{(3/2)} e^{-\Delta M_\phi/T} \right] n_\chi  \sim H(T_f) \ ,
\eeq
when $\chi$ freezes-out.

The entropy density is given as~\cite{CosmoBooks}
\beq
s(T) = \frac{2\pi^2}{45} g_{*} T^3 = 0.44 g_{*} T^3 \ ,
\label{entdenT.EQ}
\eeq
where $g_{*}$ is the number of relativistic degrees of freedom at $T$. 
Defining the ratio of $n$ to $s$
\beq
Y \equiv \frac{n}{s} \ ,
\eeq 
we find, by writing the Boltzmann equation Eq.~(\ref{sNBoltz.EQ}) in terms of $Y$, that it is 
conserved (i.e. $dY/dt = 0$) provided: (a) the $\left<\sigma v\right>$ term is negligible, 
and (b) $C_\Gamma$ is negligible\footnote{If $C_\Gamma$ is 
not negligible see Eq.~(\ref{dYdtCG.EQ}).}. 
If the above two conditions are satisfied, the conserved value after freeze-out is given by
\beq
Y_f = \frac{n_f}{s_f} = \frac{3.77}{g_{*f}} \frac{x_f}{M M_{Pl}} \frac{1}{\left< \sigma v\right>} \ , 
\eeq
where $M$ is the mass of $\chi$ and $x_f \equiv M/T_f$.
The entropy density today $s_0$ is given by
\beq
s_0 = \frac{2\pi^2}{45} g_{*0} T_0^3 \approx (3\times 10^{-4}~{\rm eV})^3 = 3678~{\rm cm}^{-3} \ ,
\label{s0T0.EQ}
\eeq 
where we have used $g_{*0} = 6.3$ and $T_0 = 2.7~{\rm K} = 2\times 10^{-4}~{\rm eV}$ is the
photon temperature today, from which we get the number density of $n$ today
\beq
n_0 = Y_f s_0 \ .
\label{n0Yfs0.EQ}
\eeq
We can then compute the present relic energy density of the species $\chi$ from
\beq
\Omega_0 \equiv \frac{n_0 M}{\rho_c} = \frac{10^{-10} x_f {\rm eV}^3}{g_{*f}^{0.5} M_{Pl}\, \rho_c \left< \sigma v\right> } \ , 
\label{Om0def.EQ}
\eeq
where $\rho_c = (2.95\times 10^{-3} \sqrt{h}~{\rm eV})^4 $ is the critical energy density.
With $g_{*f} \approx 100$, we get (cf. Eq.~(3.4) of Ref.~\cite{Jungman:1995df})
\beq
\Omega_0 h^2 = 10^{-29} x_f \left(\frac{{\rm eV}^{-2}}{\left<\sigma v\right>}\right) = 1.1\times 10^{-28} x_f \left(\frac{{\rm cm}^3/{\rm s}}{\left<\sigma v\right>}\right) \ .
\label{Omsigxf.EQ}
\eeq

For a species that is non-relativistic at freeze-out, i.e., cold dark matter (CDM),
the number density is given by (cf. Eq.~(\ref{numdenT.EQ})) 
\beq
n_f = g \left(\frac{M~T_f}{2\pi}\right)^{3/2} e^{-M/T_f} \ .
\eeq
From Eq.~(\ref{eqIntH.EQ}), $x_f$ is implicitly given by (cf. Eq.~(2) of Ref.~\cite{Griest:1990kh}) 
\beq
x_f = \ln{\left(\frac{0.038 g M M_{Pl} \left<\sigma v\right>}{g_{*f}^{0.5} x_f^{-0.5}} \right)} \ .
\label{Impxf.EQ}
\eeq
$x_f$ depends logarithmically on $M$, as can be seen from Eq.~(\ref{Impxf.EQ}).
In order to obtain the observed $\Omega_0 \approx 0.3$ (with $h^2 \approx 0.5$), 
for $M \approx 100~{\rm GeV}$, $g=1$, we therefore need $x_f \approx 21$, or equivalently,
$\left<\sigma v\right> \approx 1.5\times 10^{-9}~{\rm GeV}^{-2} \approx 0.45~{\rm pbarn}$.

\section{Mixed sneutrino thermal dark matter}
\label{ThRelSnu.APP}
We found in Sec.~\ref{CONDTH.SEC} that if $s_1 > 10^{-3}$ then the $\snulsp$ will be kept in 
thermal equilibrium by self- and co-annihilations. A thermal relic results if this condition is 
satisfied, and in this section, we compute the present relic density resulting from such a 
thermal population of $\snulsp$. This has been explored earlier in~\cite{Arkani-Hamed:2000bq}.

With the lepton-number-violating parameters nonzero the $\snulsp\snulsp Z$ coupling is 
zero~\cite{Hall:1997ah}. This prevents the too efficient $\snulsp$ self-annihilation via s-channel
$Z$-boson exchange and therefore leads to an adequate relic-density as we will show here. 
Also, due to the $s_1^2$ suppression in the $\snulsp$-nucleon interaction cross-section, 
direct-detection constraints are easily evaded~\cite{Arkani-Hamed:2000bq}. 
Given that there are three generations, the $\snulsp$ are really three states, and the presence 
of these ``co-LSP'' states may lead to interesting effects depending on mass-splittings between
them. For instance, the $\snulsp \snulsp^\prime Z$ coupling may now be non-zero 
(where the $\snulsp^\prime$ is a heavier co-LSP state), and have implications in 
direct-detection experiments. We will not explore this further in this work.

In App.~\ref{ThRDCalc.APP} we gave the details of estimating the
present relic abundance, and Eq.~(\ref{Omsigxf.EQ}) is of particular relevance here.
In principle, $\left<\sigma v\right>$ is the sum of the cross-sections of all the processes that 
we identified in Sec.~\ref{SELFINT.SUBSEC} and \ref{COINT.SUBSEC}. However, for simplicity we 
consider the case when $\tilde H, \snuH, \tilde\ell$ are all much heavier 
in which case the only processes that are relevant are $(a_s)$ and $(c_s)$.
We thus have $\left<\sigma v\right> \approx \sigma_{SA}^{(a_s)} + \sigma_{SA}^{(c_s)}$,
and for $x_f \approx 21$, the present relic abundance of $\snulsp$ is
\beq
\Omega_0 h^2 = \frac{10^{-4}}{s_1^4} 
\left\{ \left[g^2\left(\frac{100~{\rm GeV}}{M_{\tilde W}}\right) + 
g^{\prime\, 2} \left(\frac{100~{\rm GeV}}{M_{\tilde B}} \right) \right]^2 +  
\frac{Y_\psi^2}{t_1^2} \left(\frac{|A_N -\mu^* Y_N|}{M_{h}}\right)^2 
\left(\frac{100~{\rm GeV}}{M_{h}}\right)^2 \right\}^{-1} \ .
\label{Om0s1.EQ}
\eeq

If $\snulsp$ is predominantly right-handed ($s_1 \ll 1, c_1\approx 1$, $\snulsp \approx \sNR$), 
if $|A_N -\mu^* Y_N| \gtrsim M_{h}$, and if 
$\Delta M_{\snuH} \equiv (M_{\snuH} - \Mlsp)$ is not too large
compared to $T$, co-annihilation of the $\sNR$ with $\snuH$ by process $(e_c)$ 
can be effective, with the cross-section given in Table~\ref{CoIntSUSY.TAB}. 
In this case, the present relic-density is
\beq
\Omega_0 h^2 = \frac{1}{c_1^4} e^{(\Delta M_{\snuH}/T)} \left(\frac{M_{h}}{100~{\rm GeV}} \right)^2 \left( \frac{M_{h}}{|A_N -\mu^* Y_N|} \right)^2
\eeq
We can see from Eq.~(\ref{thsnu.EQ}) that in order to have $s_1 \ll 1$ when 
$|A_N -\mu^* Y_N| \gtrsim M_{h}$ 
requires $m_\ell \gg v$. This implies that $\Delta M_{\snuH}$ is necessarily large,
leading to too big a relic-density, and excluding this possibility.

In general we find that if a thermal population of 
$\sNR$ is generated due to the processes of Secs.~\ref{SELFINT.SUBSEC} and \ref{COINT.SUBSEC}, 
then the resulting relic density is too big. The reason for this is that the co-annihilation
cross-section is big enough to thermalize the $\sNR$, but is too small to keep it in thermal
equilibrium long enough to deplete the number-density to acceptable levels before freeze-out.
Therefore, if the $\snulsp$ is in thermal equilibrium we must have $s_1$ not too small, as 
dictated by Eq.~(\ref{Om0s1.EQ}).

\section{Relic from decay of thermal SUSY particles}
\label{ReThDec.APP}
At some point in the history of the universe, if there exists a thermal population of heavier SUSY 
particles, they could decay into the $\snulsp$ giving rise to a relic abundance~\cite{Asaka:2005cn}.
Here we focus on the case $s_1 \ll 1$ i.e., $\snulsp \approx \sNR$.
The important decay channels are:
\begin{itemize}
   \item[$(a_D)$] $\tilde H_u \rightarrow \snulsp L$ (where $L$ is the lepton doublet).
   \item[$(b_D)$] $\snuH \rightarrow \snulsp \overline{\psi} \psi$ via $h_u,h_d$ (where $\psi$ is a SM fermion).
   \item[$(c_D)$] $\tilde\ell \rightarrow \snulsp \overline{\psi} \psi^\prime$ via $W^\pm$. 
\end{itemize}

We find the cumulative relic abundance of $\sNR$ today by integrating the Boltzmann equation, 
Eq.~(\ref{sNBoltz.EQ}). We take as an estimate for $C_\Gamma$,
\beq
C_\Gamma \sim n_{\chi}\Gamma\left(\chi \rightarrow \snulsp X \right) \ ,
\label{CGamHino.EQ}
\eeq
where $\Gamma$ is the decay rate of the SUSY particle $\chi$ in processes $(a_D)$-$(c_D)$,
and $X$ denotes SM particles in these processes. In this estimate, we ignore the angular 
dependence in the decays.
The Boltzmann equation written in terms of $Y_{\snulsp}$ is 
\beq
\frac{d}{dt} Y = \frac{C_\Gamma}{s} \ , 
\label{dYdtCG.EQ}
\eeq
in the limit where the $\left<\sigma v\right>$ term can be neglected in Eq.~(\ref{sNBoltz.EQ}).
In the radiation dominated era we can change the independent variable from $t$ to $T$ using 
\beq
dt = -\frac{dT}{HT} \ , 
\label{dttodT.EQ}
\eeq
after which Eq.~(\ref{dYdtCG.EQ}) can be written as 
\beq
\frac{d}{dT} Y = - \frac{C_\Gamma}{H T s} \ .
\eeq
Integrating this from $T_{\infty}$ down to $T$, we have
\beq
Y(T) = -\int_{T_\infty}^{T} dT \frac{C_\Gamma}{H T s} \ .
\eeq
Using Eqs.~(\ref{CGamHino.EQ}), (\ref{HT2.EQ}), (\ref{numdenT.EQ}) and (\ref{entdenT.EQ}) 
in the above equation we get
\beq
Y(T) \sim \frac{\Gamma M_{Pl}}{M_{\chi}^2} \int_{T/M_{\chi}}^{\infty} dx \frac{e^{-1/x}}{x^{9/2}} \ .
\label{YGam.EQ}
\eeq 
Owing to the exponential suppression in the integrand for $T < M_\chi$, $Y$ freezes in at 
$Y_f \approx Y(M_\chi)$. 

The decay rate from process $(a_D)$ (in the $\tilde H$ rest-frame) is given by
\beq
\Gamma_{(a_D)} \sim \frac{c_1^2 Y_N^2}{16\pi} M_{\tilde H} f_{PS}^2(\Mlsp,M_{\tilde H}) \ , \qquad f_{PS}^2(M_x,M_y) = \frac{\left( 1-M_x^2/M_y^2\right)^2}{\left(1+M_x^2/M_y^2\right)} \ ,
\label{GamHinoLN.EQ}
\eeq
where we have taken the lepton in the final state as massless, $c_1 = \cos{\theta^{\snu}_1}$, and
$f_{PS}$ is the phase-space function.
Due to this decay, from Eq.~(\ref{YGam.EQ}), we get
\beq
Y_{(a_D)}(T) \sim 10^{-2} c_1^2 Y_N^2 f^2_{PS} \frac{M_{Pl}}{M_{\tilde H}} \int_{T/M_{\tilde H}}^{\infty} dx \frac{e^{-1/x}}{x^{9/2}} \ .
\eeq 
Using Eqs.~(\ref{s0T0.EQ}), (\ref{n0Yfs0.EQ}) and (\ref{Om0def.EQ}) we find the present relic 
density to be
\beq
\Omega_{0\,(a_D)} h^2 \sim 10^{26} c_1^2 Y_N^2 \frac{\Mlsp}{M_{\tilde H}} f_{PS}^2  \ .
\label{Om0ReDec.EQ}
\eeq 

The decay rate of process $(b_D)$ is given by
\beq
\Gamma_{(b_D)} \sim \frac{(c_1^2-s_1^2)^2 Y_\psi^2}{256\pi^3} \frac{|A_N -\mu^* Y_N|^2 M_{\snuH}^3}{M_{h}^4} f_{3PS}^2 \ ,  
\eeq
where $f_{3PS}$ is the 3-body phase-space, and following the same procedure as in the 
previous case, we find 
\beq
\Omega_{0\,(b_D)} h^2 = 10^{24} (c_1^2-s_1^2)^2 Y_\psi^2 
\frac{|A_N -\mu^* Y_N|^2 M_{\snuH} \Mlsp}{M_{h}^4} f_{3PS}^2  \ .
\eeq

The decay rate of process $(c_D)$ is given by
\beq
\Gamma_{(c_D)} \sim \frac{s_1^2 g^4}{256\pi^3} \frac{M_{\tilde\ell}^5}{M_{W}^4} f_{3PS}^2 \ ,  
\eeq
and we find 
\beq
\Omega_{0\,(c_D)} h^2 = 10^{24} s_1^2 g^4 \frac{M_{\tilde\ell}^3 \Mlsp}{M_{W}^4} f_{3PS}^2   \ .
\eeq

From these estimates, we infer that for $\Omega_0 h^2 \lesssim O(1)$, with all masses about 
100~GeV, we need: $Y_N \lesssim 10^{-13}$, $|A_N -\mu^* Y_N| \lesssim 10~$eV, and 
$s_1 \lesssim 10^{-12}$ (where we ignore possible phase-space suppression factors). 
This is realized for instance for a purely Dirac neutrino,
and has recently been considered in Ref.~\cite{Asaka:2005cn}.
In this case the heavier SUSY particles are very long lived, and one has to worry if it decays 
during big-bang nucleosynthesis (BBN) causing photo-dissociation of the light elements, and 
spoiling its successful predictions. We find from Eq.~(\ref{dttodT.EQ})
that in order for heavy particles to have decayed away before
BBN, we need the life-time of any such particle to be $\tau < 1~{\rm s}$, which implies that
the width $\Gamma > 10^{-25}~{\rm GeV}$. 
Considering for example the decay of $\tilde H$, we find from Eq.~(\ref{GamHinoLN.EQ})
that we need $Y_N > 10^{-13}$ in order for the $\tilde H$ to have decayed away before BBN.
It is worth mentioning that a scenario similar to the one explored in \cite{Asaka:2005cn} might be realizable even if the neutrinos are Majorana fermions. It seems, for example, that 
an ``eV-seesaw'' \cite{deGouvea:2005er}, which imposes $M_N\sim 1$~eV and $Y_N\sim 10^{-11}$ might satisfy the conditions discussed above. A more detailed analysis of this possibility is beyond the 
ambitions of this paper.


\end{document}